\documentclass[conference]{IEEEtran}
\IEEEoverridecommandlockouts
\usepackage{geometry}
\usepackage{cite}
\usepackage{amsmath,amssymb,amsfonts}
\usepackage{graphicx}
\usepackage{textcomp}
\usepackage{xcolor}
\usepackage{float} 
\usepackage{multirow}
\usepackage{array}
\usepackage{booktabs}
\usepackage{multirow}
\usepackage{color}
\usepackage{nomencl}
\makenomenclature
\usepackage{etoolbox}
\usepackage[linesnumbered, ruled, vlined]{algorithm2e}
\usepackage{nicematrix}
\usepackage{mathtools}
\usepackage{booktabs}
\usepackage[caption=false]{subfig}

\usepackage[pscoord]{eso-pic}

\begin{document}

\title{SkyNetPredictor: Network Performance Prediction in Avionic Communication using AI\\}

\author{\IEEEauthorblockN{Hind Mukhtar}
\IEEEauthorblockA{\textit{Satcom Direct, Inc.} \\
\textit{Melbourne, Florida}\\
\textit{University of Ottawa}\\
\textit{Ottawa, Ontario}\\
hmukhtar@satcomdirect.com\\
hmukh045@uottawa.ca}
\and
\IEEEauthorblockN{Raymond Schaub III}
\IEEEauthorblockA{
\textit{Dept. of Data Science}\\
\textit{Satcom Direct, Inc.} \\
\textit{Melbourne, Florida}\\
rschaub@satcomdirect.com}
\and
\IEEEauthorblockN{Melike Erol-Kantarci}
\IEEEauthorblockA{
\textit{IEEE Fellow}\\
\textit{Dept. of Electrical \&}\\
\textit{Computer Engineering}\\
\textit{University of Ottawa }\\
\textit{Ottawa, Ontario}\\
melike.erolkantarci@uottawa.ca}
}


\maketitle

\begin{abstract}
Satellite-based communication systems are integral to delivering high-speed data services in aviation, particularly for business aviation operations requiring global connectivity. These systems, however, are challenged by a multitude of interdependent factors such as satellite handovers, congestion, flight maneuvers and seasonal trends, making network performance prediction a complex task. No established methodologies currently exist for network performance prediction in avionic communication systems. This paper addresses the gap by proposing machine learning (ML)-based approaches for pre-flight network performance predictions. The proposed models predict performance along a given flight path, taking as input positional and network-related information and outputting the predicted performance for each position. In business aviation, flight crews typically have multiple flight plans to choose from for each city pair, allowing them to select the most optimal option. This approach enables proactive decision-making, such as selecting optimal flight paths prior to departure. 
\end{abstract}


\section{Introduction}\label{Introduction}
The increasing reliance on high-speed data services for aviation has spurred the need for robust and efficient communication systems capable of supporting diverse operational and passenger demands \cite{IATA}. Satellite-based communication systems have emerged as a vital solution for avionic communication, offering global coverage and high data throughput, even in remote or transoceanic regions \cite{10328056}. However, ensuring optimal performance in such systems is a complex challenge due to multiple interdependent factors such as the flight path, satellite beam availability, signal-to-noise ratio (SNR), Maximum Information Rate (MIR), congestion, flight maneuvers, weather conditions, seasonal trends, and fluctuating user demands. This challenge is especially significant in business aviation, where unpredictable traffic patterns and fluctuating user demands add further complexity compared to commercial flights. 

Currently, network performance prediction for avionic communication lacks established methodologies, largely due to the complexity and variability of satellite-based communication systems.  This lack of established approaches necessitates the exploration of ML techniques, which offer the ability to model complex dependencies, learn patterns from diverse datasets and adapt to dynamic conditions. Our previous work in \cite{10625144} investigated the use of ML techniques for air traffic value and user demand predictions within a satellite beam. That study demonstrated that Variational Auto-encoder Generative Adversarial Networks (VAEGANs) achieved good prediction performance and were capable of capturing sudden fluctuations in trends for both high-traffic and low-traffic beams. However, a limitation of this approach was that the number of output nodes in the model increased with the number of beams, making it difficult to scale for real-world applications.

In this paper, we build upon that work by proposing the use of a Long Short Term Memory (LSTM) model to predict network performance along a given flight path, irrespective of the number of beams. The model takes positional and network-related information as input and outputs the predicted performance for each position along the path, addressing scalability challenges. This approach enables proactive decision-making, such as selecting a flight path with optimal performance, before the flight begins, by leveraging the LSTM model's ability to capture temporal dependencies and predict network performance at various flight positions. Additionally, we extend this problem to use a K-Nearest Neighbors (KNN) model. We also benchmark our LSTM and KNN based approaches against a non-ML, rule-based network performance prediction method to provide a comprehensive evaluation. We use a real-world dataset derived from a geostationary Earth orbit (GEO) network in the business aviation sector, leveraging historical and real-time data on flight paths, signal quality, satellite beam metrics, and user demands. Through the use of ML techniques, this paper aims to demonstrate the potential of data-driven approaches to revolutionize satellite-based communication systems in aviation. Our contributions are summarized below: 

\begin{itemize}
  \item We design a scoring system to evaluate network performance, with scores from 1 to 10, where 1 represents the poorest performance and 10 indicates the best
  \item We use a clustering algorithm to determine the probability of handovers in various regions   
  \item We propose the use of ML models for predicting network performance along a specified flight path prior to the flight, namely LSTM and KNN models 
  \item We benchmark our models against a rule-based approach to validate their effectiveness 
\end{itemize}

\section{Related Work}\label{Related Work}
The use of ML in avionic communication remains a largely underexplored research area. To our knowledge, no existing work directly addresses the specific challenge of performance prediction in satellite-based avionic communication systems. However, to establish a foundation for innovative solutions and to highlight the research gap, we focus on prior work in satellite communication, particularly from a satellite perspective rather than that of an aircraft terminal. We provides an overview of ML-based solutions for challenges faced in satellite communication systems to offer a pathway to extend these concepts to avionic communication. 

We begin our literature review by introducing the work from \cite{9622204}, as it provides a comprehensive overview of artificial intelligence (AI) applications in satellite communication. The authors establish a strong foundation by extensively covering a wide range of challenges, such as resource management, interference mitigation, energy efficiency, beam hopping, anti-jamming, traffic forecasting, telemetry mining, and channel modeling. They demonstrate how various AI techniques, including supervised learning, reinforcement learning, and deep learning, address these challenges by optimizing resource allocation, enhancing adaptability, and improving reliability in satellite networks. Moreover, the paper emphasizes the unique difficulties faced by satellite systems, such as limited onboard resources, high mobility, and frequent handoffs, and connects these to innovative AI-driven solutions. By reviewing performance metrics and results, the authors highlight tangible benefits such as reduced operational complexity and improved efficiency. 

Building on this foundational review, the study presented in \cite{10486541} explores the application of AI for enhancing handover management in hybrid satellite-terrestrial networks. This work addresses challenges posed by dynamic environments and diverse connectivity requirements, offering a system that integrates deep learning, reinforcement learning, and machine learning techniques to optimize handover processes. Their proposed predictive handover mechanism analyzes user mobility and network conditions to preemptively initiate transitions, reducing dropped connections and improving user experience. Additionally, multi-criteria decision-making frameworks are employed to balance factors such as signal strength, latency, bandwidth, and network load. The study demonstrates notable improvements in network reliability, reduced handover delays, and enhanced throughput, especially in scenarios involving high mobility or sparse terrestrial infrastructure. 

Expanding the scope to interference management, the paper \cite{8815532} introduces a deep learning-based framework for managing interference in hybrid terrestrial and satellite communication systems. This work presents two subsystems: an interference detector leveraging deep neural network autoencoders and an interference classifier utilizing LSTM networks. The detector identifies interference by analyzing deviations in input-output similarity, while the classifier distinguishes between interference types from cellular standards. The results illustrate the framework's high detection accuracy and effective classification across varying signal-to-interference ratios (SIR). This study highlights the potential of AI-driven approaches for efficient and scalable interference management in hybrid communication networks.

In summary, the prior works reviewed demonstrate the potential of ML to address critical challenges in satellite communication, such as resource allocation, interference mitigation, and handover management. These studies highlight the impact of ML techniques, including deep learning and reinforcement learning on optimizing satellite network performance. While these advancements focus primarily on satellite communication from a satellite-centric perspective, our work extends these methodologies into the domain of avionic communication systems. By leveraging real-world flight data, our proposed ML-based network performance prediction framework aims to address the unique challenges of satellite-based avionic communication. This approach has the potential to improve network reliability and enhance user experience. 

\section{Dataset}\label{Dataset}
In this paper, we use a real-world dataset containing anonymized positional and network performance information for business aviation jets. The dataset includes timestamps, positional data, satellite and beam identifiers, handover probabilities, historical averages for SNR and MIR and the user’s historical average flight performance and usage trends. The data spans flights to and from the United States from January to October 2024 and pertain to a GEO satellite constellation operating in the Ku-band. Before inputting the data into the model, several preprocessing and augmentation steps are performed. Each preprocessing step is described in detail below.

\subsection{Handover Probability by Position}
To calculate handover probabilities, we begin by retrieving historical handover and positional data. Handovers are categorized into three distinct types: satellite handovers, make-before-break beam handovers, and break-before-make beam handovers. This classification is critical because each type affects network performance differently. For example, satellite handovers typically take longer than beam handovers, while break-before-make beam handovers—indicating a switch between beams of opposite polarization—require extended switch times \cite{Gannon_Schoenholz_Downey_Clapham_2024a}. A break-before-make handover occurs when the current connection is terminated before establishing a new one. This type of handover typically happens during a switch between beams with opposite polarizations or when transitioning between beams with non-overlapping coverage areas. In such scenarios, the existing link is severed because maintaining both connections simultaneously is not technically feasible \cite{prasad2003applied}. Additionally, handover data is segmented by heading (Northwest, Northeast, Southwest, Southeast) to ensure that handover probabilities reflect directional consistency. This prevents high handover probabilities from being erroneously triggered for aircrafts traveling away from a beam. 

We then apply DBScan, to identify handover regions based on density and generate geospatial polygons within these clusters. To refine the clustering, we perform DBScan again within each cluster, reducing the minimum distance between points to break large clusters into smaller, higher-density clusters. This process is detailed in Algorithm \ref{alg: DBScan}, where $num\_layers$ represents the number of contour layers desired, $min\_distance$ specifies the smallest $\epsilon$ value used in DBScan for the densest layer, and $max\_distance$ defines the largest $\epsilon$ value used for the outermost layer. Here, $\epsilon$ is a key parameter in DBScan that determines the maximum distance between two points for them to be considered part of the same cluster. Once the contoured clusters are generated, we compute the handover probability for each cluster. This is done by dividing the number of handovers within each child cluster by the total number of handovers in the parent cluster. This hierarchical approach provides a nuanced and directionally aware estimation of handover probabilities across varying regions. 

\begin{algorithm}[!t]
\caption{Generate Contoured Handover Regions}
\KwData{$num\_layers$, $min and max distance$, handover data}
\KwResult{Contoured Handover Regions}

\textbf{Step 1: Compute Step Size}\\
$step = \frac{max\_distance - min\_distance}{num\_layers}$\;

\textbf{Step 2: Initialize Clustering}\\
Initialize clustering parameters $\epsilon_1$, $min\_samples$ and set parents clusters using DBSCAN\;

Generate polygons for parent clusters \;

\textbf{Step 3: Generate child clusters}\\
\For{each layer $i = 0$ to $num\_layers - 1$}{
    $eps2$ = $max(esp1 - step, min\_distance)$\;
    
    Assign child clusters using DBSCAN given $\epsilon_2$\;

    Generate polygons for child clusters
    
    Subtract inner polygons from current polygons\;
    
    Update cluster IDs and parent clusters for new polygons\;
}
\label{alg: DBScan}
\end{algorithm}

\subsection{Historical Performance}
To determine the historical SNR and MIR by location, we begin by dividing the footprint of each beam into smaller hexagons, each approximately with a radius of 50 km. Then, we calculate the average historical SNR and MIR for each of these smaller hexagons, using historical flight data to compute the averages for every hexagons within the beam's footprint. This approach allows us to assess the historical network performance across different areas of the beam, which is important as performance can vary within a beam due to factors such as the beam pattern, interference from neighboring beams, and other environmental influences.

For each flight, we calculate the aircraft's average network performance score over the past 5 flights, as well as the average number of connected devices during the same period. This provides insights into the state of the avionic communication system installed on the aircraft. For example, if the aircraft has experienced poor performance on average in its previous flights, it may indicate an anomaly in the system rather than issues with the network itself. Similarly, if the aircraft typically has a high number of connected devices, it increases the likelihood of experiencing congestion due to higher demand.

\subsection{Augmented Time and Position Data}
The flight data is provided at 30-second intervals. However, to reduce the frequency of predictions, we sample the data at 10-minute intervals. By making predictions at 10-minute intervals, we gain insight into where a flight might experience good or poor network performance along its path, which is sufficient for our purpose. This simplification reduces the model complexity. Additionally, we use the timestamps in the flight data to extract the year, month, day, hour, and minute. We also include a season index to indicate the season, a holiday index to determine whether the day is a holiday, and a weekend index to classify whether the day is a weekday or weekend. This method, applied in our first paper, \cite{10625144}, proved effective in identifying seasonal trends related to air traffic and user demand. To enhance the positional data, we transform the longitude and latitude by taking their sine and cosine to capture their cyclical nature. We also calculate the aircraft's heading at each timestamp along the flight path. Additionally, we compute the time and distance since the flight's start, as well as the time and distance remaining to the destination.

\subsection{Sequencing, Data Normalization and Train-Test Split}
After completing the augmentation and preprocessing steps described above, we generate input and output sequences for each flight. Each flight is treated as a sequence, where the input features form the input sequence and the network performance score at a given time constitutes the output sequence. The network performance score is an integer between 1 and 10, where 1 represents the poorest performance and 10 represents the best performance. This score is determined based on a combination of terminal status, SNR, and demand allocation. Since predictions are made prior to the flight, the output is never used as input to the model. The goal of the model is to learn the mapping from the input sequence to the output sequence. The input and output features are summarized as follows:
\begin{itemize}
    \item \textbf{Input Features:} Year, month, day, hour, minute, season, holiday and weekend index, aircraft tail number, destination and departure airport, longitude sin and cosine, latitude sin and cosine, altitude, heading, distance and time since start, distance and time to destination, average SNR, MIR, score and number of connected devices, satellite and beam name, satellite, make before break and break before make handover probability
    \item \textbf{Output Features:} Network Performance Score
\end{itemize}

We treat this as a classification problem rather than a regression problem. Since the network performance score is an integer between 1 and 10, there are 10 output classes, each representing a specific score. Consequently, we convert the output sequence into a one-hot encoded value. We apply min-max normalization to the input data and then separate the dataset into training, validation, and testing sets by shuffling the sequences with a ratio of 0.8 for training, 0.1 for validation, and 0.1 for testing.

\section{Proposed Technique}\label{Proposed Technique}
In this paper, we propose LSTM and KNN models to predict the network performance score of a satellite-based avionic communication system along a given flight path. The proposed models are optimized to make predictions prior to the flight, meaning they do not depend on outputs from previous timestamps for subsequent predictions. Additionally, the LSTM model is designed to handle varying input and output sequence lengths, making it adaptable to flights of different durations. To evaluate and compare the performance of our proposed models, we also deploy a non-ML rule-based approach. Details of the proposed and benchmark models are provided below.

\subsection{LSTM Model Architecture}
The LSTM model processes 36 input features of varying sequence lengths and outputs probabilities for 10 classes. It consists of 4 LSTM layers, each with a hidden size of 128, followed by a fully connected layer of size 128 and an output layer of size 10. A dropout layer is applied after the LSTM output to prevent overfitting \cite{10.5555/2627435.2670313}. The softmax activation function is used at the output layer to generate class probabilities. To accommodate variable-length input and output sequences, the model pads sequences to match the length of the longest sequence within each batch. The use of masks during training and inference prevents the model from considering the padded values. We use a weighted cross-entropy loss function to address the class imbalance in the dataset, where higher weights are assigned to lower scores. This approach compensates for the natural bias of the dataset, as the network generally performs well, leading to a predominance of better-performing scores. The weights are inversely proportional to the frequency of each class in the training data, ensuring that rare, low scores contribute more significantly to the loss function. This technique ensures that the model learns to predict poor network performance more accurately. 

For training, we use the Adam optimizer with a weight decay factor of 0.0001 to regularize the model and prevent overfitting. We start with an initial learning rate of 0.001 and employ a Reduce-On-Plateau learning rate scheduler, which reduces the learning rate by a factor of 10\% when the validation loss plateaus for 5 epochs. This ensures a smooth and adaptive optimization process. The model is trained with a batch size of 128, and we implement gradient accumulation with an accumulation factor of 4, resulting in an effective batch size of 512. Gradient accumulation allows us to simulate a larger batch size by summing gradients across multiple mini-batches before performing an optimization step, enabling more stable training without exceeding memory constraints \cite{bhattacharyya_2020}. Additionally, we employ early stopping to monitor the validation loss. The model is implemented using pytorch \cite{paszke2017automatic}

\begin{figure}[t] 
    \centering
    \includegraphics[scale = 0.5]{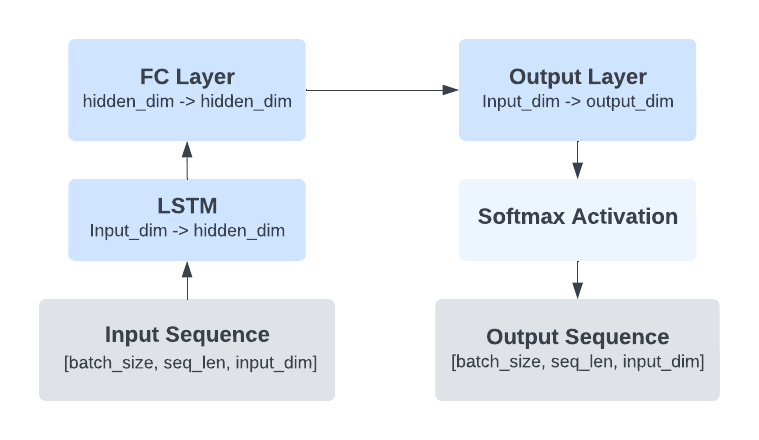} 
    \caption{LSTM model block diagram}
    \label{fig:LSTM}
\end{figure}

\subsection{KNN Model Architecture}
The KNN model takes the same 36 input features as the LSTM model and outputs probabilities for the 10 classes representing the network performance score. The model assigns a class to an input based on the majority class among its 3 closest neighbors in the feature space. We use scikit-learn's implementation of KTree from the work in \cite{7898482}, which proposes finding the k nearest values by constructing a decision tree, to enhance the efficiency of the search algorithm. The model is trained using the training dataset by fitting the input features and their corresponding output classes. 

\subsection{Rule-Based Approach}
While no industry-standard mechanism currently exists for network performance predictions, some industry practices include real-time tracking and historical network performance statistics and analysis \cite{satcomdirect_sdpro}. The rule-based approach is intended to quantify the accuracy of such methods and serves as a benchmark model for comparison with the LSTM and KNN models. This approach uses historical averages and rules to determine the network performance score for a given flight path. The rule-based process begins by checking the location of the aircraft and associating it with a square within the beam, as mentioned in section \ref{Dataset}. The average historical performance score within each of these squares is calculated and then mapped to the input using a mapping function or lookup table. If that area has not been previously traversed, it is simply assigned the average historical performance score of the entire dataset. This simple approach provides a baseline for understanding the impact of the input features on network performance and serves as a point of comparison for more complex machine learning models.

\section{Performance Evaluation}\label{Performance Evaluation}
To evaluate the effectiveness of our proposed models, we use the dataset outlined in Section \ref{Dataset}. Additionally, we compare their performance against the rule-based approach. We consider a variety of metrics to assess the performance of all models. 


Table \ref{table: performance summary} summarizes the precision, recall, f1-score, accuracy and rmse of the proposed and benchmark models. Since this is not truly a classification task but rather a task where each output class represents an integer score between 1 and 10, we use also Root Mean Squared Error (RMSE) to evaluate the magnitude of prediction errors. In this case, accuracy does not fully capture model performance, as it assumes a prediction is entirely correct or wrong. For example, if the model predicts a score of 8 but the ground truth is 9, accuracy would count this as a wrong prediction, even though the model's prediction is very close. RMSE, however, accounts for the difference between predicted and actual values, penalizing larger errors and providing a more informative measure of how well the model approximates the true values, especially when the predicted scores are close but not exact.

Our results show that the rule-based approach exhibits the highest RMSE and the lowest values for precision, recall, F1-score, and accuracy, suggesting that a more complex model is needed to capture the intricacies of the dataset. The KNN model achieves the best performance overall, followed closely by the LSTM model. However, it is important to note that the superior performance of the KNN model comes at the cost of significantly longer inference times. While the LSTM model can predict the network performance of a given flight path within seconds, the KNN model requires several minutes to make the same prediction, highlighting a trade-off between prediction accuracy and computational efficiency. To be exact, predicting the performance of approximately 20,000 flights in the test dataset takes 22 seconds with the LSTM model, whereas the KNN model requires 67 minutes. This is because the KNN model calculates the distance between the query point and every point in the training dataset. As the dataset grows, the computational cost increases, making it difficult to scale. The dataset used for the performance evaluation only contains flights to and from the United States of America. Hence, the inference time is expected to worsen once deployed for real-world applications with global flight data. While the problem is not time-sensitive since we are performing offline predictions, the increased inference time can still be an inconvenience. This makes the LSTM model a more attractive solution, as it handles large datasets more efficiently.

\begin{table}[t]
\caption{Table summarizing proposed and benchmark model performance}
\label{table: performance summary}
\begin{tabular}{@{}llllll@{}}
\toprule
\textbf{Model}      & \textbf{Precision} & \textbf{Recall} & \textbf{F1 Score} & \textbf{Accuracy} & \textbf{RMSE} \\ \midrule
\textbf{LSTM}       & 57.6               & 61.0            & 57.3              & 61.0              & 2.34          \\
\textbf{KNN}        & 62.3               & 63.8            & 62.7              & 63.8              & 1.87          \\
\textbf{Rule-Based} & 45.5               & 11.1            & 5.8              & 11.1              & 2.32          \\ \bottomrule
\end{tabular}
\end{table}

\begin{figure*}[t] 
    \centering
    \includegraphics[scale = 0.85]{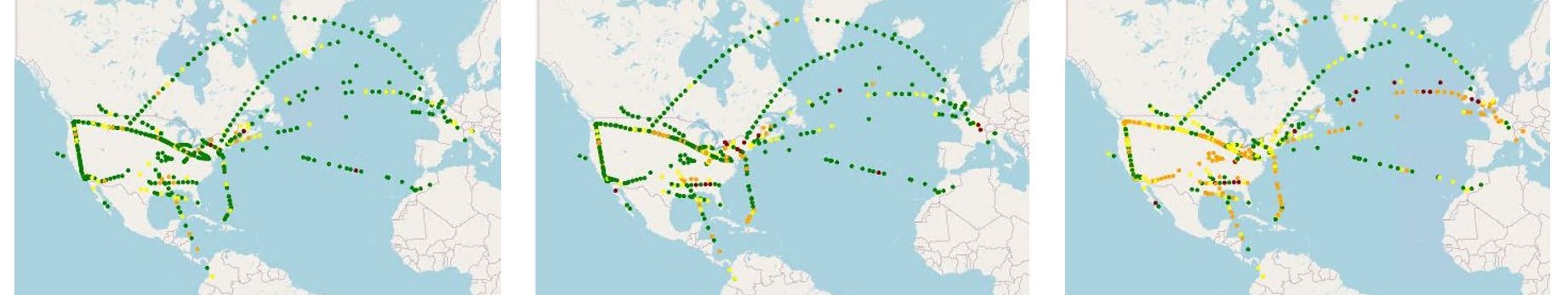} 
    \caption{Plots displaying accuracy of model predictions. Left: KNN. Center: LSTM. Right: Rule Based Approach}
    \label{fig:plot results}
\end{figure*}

Figure \ref{fig:plot results} illustrates the predicted performance scores for 20 randomly sampled flights, with a colorscale indicating the intensity of the difference between true and predicted scores. Green represents differences of less than 1, indicating high prediction accuracy, while yellow covers differences between 2 and 3, orange corresponds to differences between 4 and 7, and red signifies differences greater than 7. These results demonstrate that both the LSTM and KNN models significantly outperform the rule-based approach. Specifically, 79\% of the KNN model’s predictions are within 1 point of the true value, compared to 74\% for the LSTM model and only 54\% for the rule-based approach.

\begin{figure}[t] 
    \centering
    \includegraphics[scale = 0.8]{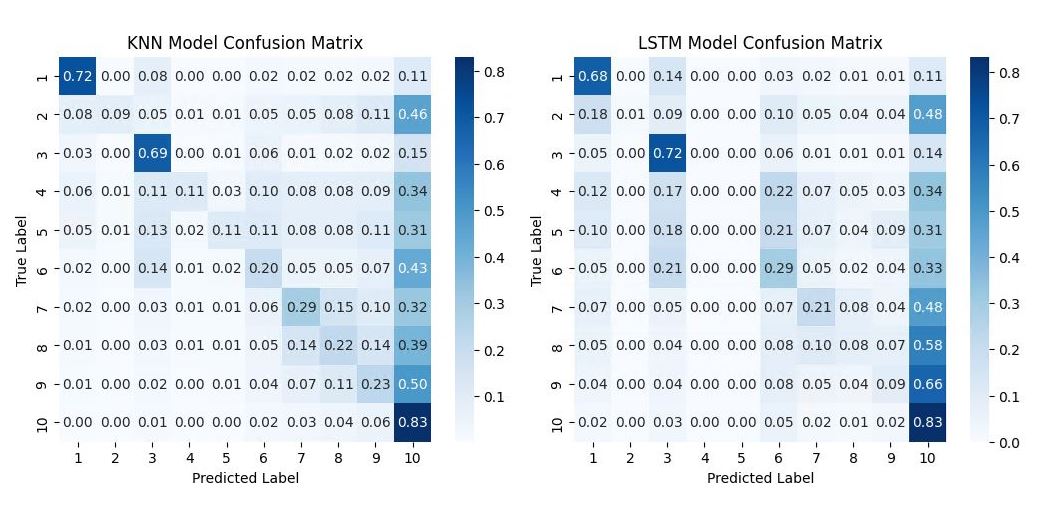} 
    \caption{Confusion Matrices for KNN and LSTM models}
    \label{fig:CM}
\end{figure}

Figure \ref{fig:CM} displays the confusion matrices for the KNN and LSTM models, showing that while both models perform best for class 10 and exhibit relatively good performance for classes 1 and 3, they struggle with the underrepresented classes, frequently misclassifying them as class 10. The KNN model seems to have slightly better performance than the LSTM model for these underrepresented classes. However, the KNN model is essentially at its optimal performance, as improvements from hyperparameter tuning are limited, particularly given the size and high dimensionality of the dataset. On the other hand, the LSTM model still has room for improvement. By exploring alternative loss functions that better address the class imbalance we could potentially enhance the model's ability to better distinguish between the underrepresented classes. The LSTM model can also be integrated into more complex deep learning architectures, such as those using attention mechanisms. This allows for future improvements in model performance and adaptability without the need to fully replace the foundational LSTM model. While the work in this paper focuses on the use of ML models for pre-flight network performance predictions, the LSTM model can easily be adapted for real-time predictions. The reduced inference time of the LSTM model, coupled with these improvements, highlights its potential for efficient network performance predictions. 

\begin{figure}[t] 
    \centering
    \includegraphics[scale = 0.9]{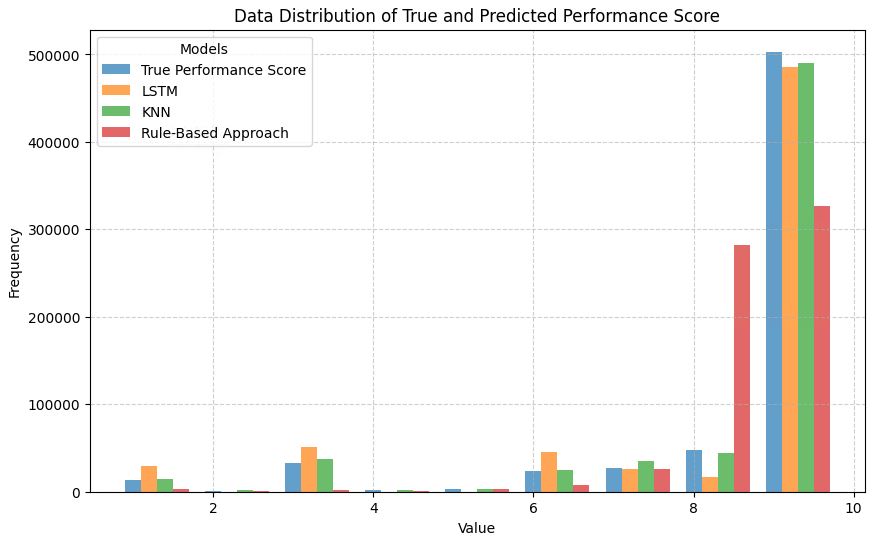} 
    \caption{Plot displaying the distribution of the true and predicted classes}
    \label{fig:data distribution}
\end{figure}

\begin{figure}[t] 
    \centering
    \includegraphics[scale = 0.85]{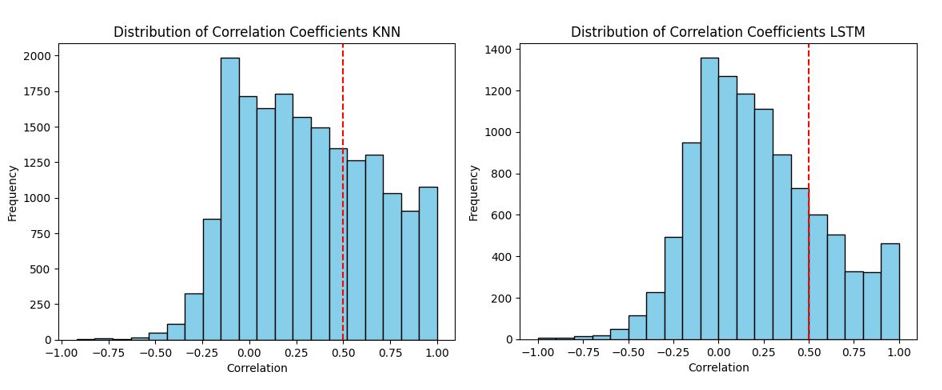} 
    \caption{Distribution of correlation coefficients}
    \label{fig:correlation}
\end{figure}

Figure \ref{fig:correlation} shows the distribution of the correlation coefficient between the predicted and true sequences. The distribution of its correlation coefficients demonstrates that the models are capable of capturing the overall trends of the sequence.

\section{Conclusion}\label{Conclusion}
In business aviation, having prior knowledge of network performance along a given flight path enables proactive decision-making, such as selecting optimal flight paths prior to departure. In this paper, we introduce LSTM and KNN based models that take historical network performance and positional data as input and output a predicted network performance score ranging from 1 to 10. Additionally, we benchmark our approach against a rule-based model. Our results show that both the LSTM and KNN models outperform the rule-based approach, with the KNN model achieving slightly better performance than the LSTM model. However, the LSTM model demonstrates a much lower inference time compared to the KNN model.

We observe a significant class imbalance, as the satellite network typically tends to perform well, leading to a large portion of samples belonging to the highest performance score. Both the LSTM and KNN models were able to make good predictions for some of the underrepresented classes, but they struggled with the most underrepresented ones. While the KNN model appears to have reached its optimal performance, with limited improvements from hyperparameter tuning particularly given the size and high dimensionality of the dataset, the LSTM model still offers room for improvement. In future work, we plan to explore alternative loss functions that better address class imbalance and experiment with more complex deep learning architectures to further enhance the model's ability to distinguish between the underrepresented classes.

\bibliographystyle{IEEEtran}
\bibliography{references} 

\end{document}